\documentclass[conference]{IEEEtran}
\IEEEoverridecommandlockouts
\usepackage{authblk}
\usepackage{cite}
\usepackage[T1]{fontenc}
\usepackage{amsmath,amssymb,amsfonts}
\usepackage{algorithmic}
\usepackage{graphicx}
\usepackage{textcomp}
\usepackage[dvipsnames]{xcolor}
\definecolor{mauve}{rgb}{0, 0, 0} 
\definecolor{azure(colorwheel)}{rgb}{0.0, 0.5, 1.0}
\definecolor{fragile-points}{HTML}{4285f4}
\definecolor{sensitive-points}{HTML}{ff6d00}
\definecolor{immunized-points}{HTML}{0f9d58}
\def\BibTeX{{\rm B\kern-.05em{\sc i\kern-.025em b}\kern-.08em
    T\kern-.1667em\lower.7ex\hbox{E}\kern-.125emX}}
\usepackage[colorlinks,linkcolor=blue,citecolor=blue]{hyperref}
\usepackage{booktabs}
\usepackage[framemethod=tikz]{mdframed}
\usepackage{algorithm}
\usepackage{algorithmic}

\usepackage{pgfplots}
\pgfplotsset{compat=1.14}
\usetikzlibrary{patterns}
\usepackage{listings}
\usepackage{amsthm}
\theoremstyle{definition}
\newtheorem{definition}{Definition}

\usepackage{array, etoolbox}
\newcounter{rowcount}
\AtBeginDocument{%
}

\mdfdefinestyle{mpdframe}{
    frametitlebackgroundcolor   =black!15,
    frametitlerule              =true,
    roundcorner                 =1pt,
    middlelinewidth             =1pt,
    innermargin                 =0.1cm,
    outermargin                 =0.1cm,
    innerleftmargin             =0.1cm,
    innerrightmargin            =0.1cm,
    innertopmargin              =0.1cm,
    innerbottommargin           =0.1cm
}

\usepackage{fancyhdr}
\usepackage{xspace}
\newcommand\tripleagent{{\sc TripleAgent}\xspace}
\newcommand\tripleagentbf{{\sc \textbf{TripleAgent}}\xspace}

\newcommand{\commentmartin}[1]{\textcolor{blue}{martin: #1}}\newcommand\mm\commentmartin

\begin{document}
\title{\tripleagent: Monitoring, Perturbation and Failure-obliviousness for Automated Resilience Improvement in Java Applications}


\author[1]{Long Zhang and Martin Monperrus}
\affil[1]{KTH Royal Institute of Technology, Sweden}
\date{} 


\maketitle

\thispagestyle{fancy}
\begin{abstract}
In this paper, we present a novel resilience improvement system for Java applications. The unique feature of this system is to combine automated monitoring, automated perturbation injection, and automated resilience improvement. The latter is achieved thanks to the failure-oblivious computing, a concept introduced in 2004 by Rinard and colleagues. We design and implement the system as agents for the Java virtual machine. We evaluate the system on two real-world applications: a file transfer client and an email server. Our results show that it is possible to automatically improve the resilience of Java applications with respect to uncaught or mishandled exceptions.
\end{abstract}


\section{Introduction}\label{sec:introduction}

In modern software, resilience capabilities are engineered through error-handling code, in particular exception handling code in managed languages such as Java and C\#.
This resilience capability is manually engineered by developers, who write the error-handling code. For example, part of their coding activity is to write try-catch blocks to handle exceptions. The problem is that exception-handling code is notably hard to write and to test \cite{Felipe:An_exploratory_study_on_exception_handling_bugs_in_Java_programs}. As a result, there often exists corner-cases where resilience is not provided by developer-written code. In production, when those corner cases are activated, the software system may simply stop providing its function because it crashes after an unhandled exception \cite{DingYuan:Simple_Testing_Can_Prevent_Most_Critical_Failures}.

In order to improve error handling, two kinds of techniques are being researched: fault injection~\cite{Natella:Assessing_Dependability_with_Software_Fault_Injection_A_Survey,Ziade:A_Survey_on_Fault_Injection_Techniques} and failure-oblivious computing~\cite{rinard2004enhancing}. Fault injection is about injecting failures to trigger a system's error handling code and to analyze the abnormal behaviour\cite{Silva:A_view_on_the_past_and_future_of_fault_injection_2013}. Failure-oblivious computing is about adding fully generic error-handling code with automated code transformation~\cite{rinard2004enhancing}. In the context of exceptions, failure-oblivious computing means automatically adding catch blocks with a default exception-handling strategy~\cite{Durieux:Exhaustive_Exploration_of_the_Failure-Oblivious_Computing_Search_Space}

In this paper, \emph{our goal is to automatically improve the exception-handling code of software applications}. This is made by first finding weaknesses in resilience and then instrumenting the application with automated exception handling.
To achieve our goal of automatically improving resilience, we design a novel system, called \tripleagent, made of three components, called ``agent'' in this paper\footnote{Here, an agent refers to the Java terminology, where it is a component that is attached to the Java Virtual Machine \cite{Ghosh:Bytecode_Fault_Injection_for_Java_Software}.}.
Those three agents, automated monitoring, automated perturbation injection, and automated failure-oblivious method validation \cite{rinard2004enhancing} are orchestrated by an agent controller. The controller analyzes all the monitored data and reveals both weaknesses and suggested improvements in the resilience capabilities.


To the best of our knowledge, \tripleagent is the first system which actively injects exceptions during execution in order to, after analysis, automatically detect failure-oblivious methods.

We evaluate \tripleagent on two real-world Java applications. One is TTorrent, a file transfer client which implements the BitTorrent protocol. The other one is HedWig, an email server. In both cases, we consider a production workload: respectively downloading a large file from the Internet, and sending and receiving emails from the server.
By applying \tripleagent, we observe that exceptions thrown from $257$(21\%) perturbation points do not lead to failures anymore, which shows an automatic resilience improvement.

To sum up, our contributions are the following.

\begin{itemize}
\item The concept of joint usage of fault injection and failure-oblivious code instrumentation to evaluate and improve resilience against uncaught or mishandled exceptions. We propose a corresponding novel algorithm for automatic improvement of software resilience.

\item A system called \tripleagent that combines monitoring, perturbation injection and failure-oblivious computing in Java, implemented with agents for the Java Virtual Machine.
The system is publicly-available for future research in this area at \url{http://bit.ly/tripleagent-repo}.

\item An empirical evaluation of \tripleagent on two real-world applications of $20.3K$ lines of code in total. By performing $9968$ fault injection experiments under a realistic, production-like workload, it shows \tripleagent's effectiveness for improving software resilience. 
\end{itemize}

The rest of the paper is organized as follows: \autoref{sec:background} introduces the background. \autoref{sec:theDesignOfTripleAgent} and \autoref{sec:evaluation} present the design and evaluation of \tripleagent. \autoref{sec:relatedWork} discusses the related work, and \autoref{sec:conclusion} summarizes the paper.

\section{Background}\label{sec:background}
\tripleagent is founded on techniques from the fault injection and failure-oblivious computing \cite{rinard2004enhancing}. This section presents a basic introduction to the core concepts.

\subsection{Fault Injection}
Fault injection is a popular research topic in software testing and dependability evaluation. Fault injection techniques actively inject different kinds of errors into a target system in order to assess its dependability\cite{Natella:Assessing_Dependability_with_Software_Fault_Injection_A_Survey,Ziade:A_Survey_on_Fault_Injection_Techniques,Silva:A_view_on_the_past_and_future_of_fault_injection_2013}. This can happen in several phases:
1) during unit testing, fault injection generates more test cases so that corner cases are detected, and the coverage of testing is improved.
2) during integration, fault injection can trigger different failure scenarios so that developers gain more confidence in their system's error-handling design.
3) when done in production, fault injection is usually called ``chaos engineering'' \cite{Basiri:Chaos_Engineering:IEEESoftware}.

The kinds of failures that can be injected vary depending on the considered dependability aspect. For example, injecting processor errors or hardware-based errors is often done for evaluating the dependability of operating systems \cite{Seungjae:DOCTOR,Parasyris:GemFI_A_Fault_Injection_Tool_for_Studying_the_Behavior_of_Applications_on_Unreliable_Substrates:DSN}. Injecting connection errors between different micro-services supports to test a service's retry logic and its robustness of interacting with other services\cite{Chang:Chaos_Monkey_Increasing_SDN_Reliability_Through_Systematic_Network_Destruction}. Injecting an exception in a certain method is useful for validating an application's error-handling capability\cite{Zhang:Amplifying_Tests_to_Validate_Exception_Handling_Code}.

In this paper, we focus on fault injection in the context of Java applications, which means rather high-level, application-level fault injection.

\subsection{Failure-oblivious Computing}
In order to improve the resilience of an application, different techniques can be applied to prevent the application from crashing when an error occurs\cite{Monperrus:Automatic_Software_Repair_A_Bibliography}. Failure-oblivious computing \cite{rinard2004enhancing} is one of these approaches to overcome software failures at runtime. The main idea of failure-oblivious computing is to discard certain failures in a principled way. For example, if a method tries to write data into an invalid memory address, with failure-oblivious computing, the writing operation would be ignored. It has been shown that failure-oblivious computing is able to increase availability \cite{rinard2004enhancing,Durieux:Exhaustive_Exploration_of_the_Failure-Oblivious_Computing_Search_Space,Rigger:Lenient_Execution_of_C_on_a_Java_Virtual_Machine}, eg to serve requests to more users despite errors.

In this paper, we use the concept of failure-oblivious computing for the Java programming language. In Java, there is no invalid memory addresses, but the biggest reason for crashing are exceptions thrown at runtime. Thus, we do failure-oblivious computing for uncaught and mishandled exceptions.


\section{Design of \tripleagent}\label{sec:theDesignOfTripleAgent}
This section presents the design of \tripleagent, including relevant definitions, algorithms and its architecture.

\subsection{Definitions}\label{sec:definitions}

\paragraph{Exception} All major programming languages provide a way to signal problems through so-called exceptions. In some statically typed languages such as  C\# and Java, exceptions are typed, and some of them are statically verified at compile-time (e.g., checked exceptions in Java). For these checked exceptions, developers need to either handle them at the call site or explicitly declare them in the method signature (with the keyword $throws$ in Java\cite{Gosling:The_Java_Language_Specification_Java_SE_8_Edition})

\paragraph{Perturbation point}
In this paper, a ``perturbation point'' is a unique location in code where a fault can be injected. In \tripleagent, the considered perturbations are injected exceptions, which means that a perturbation point is defined as a statement that potentially throws an exception. A perturbation point is noted $<m, l, e>$. $m$ describes the method where this point is located. $l$ is the line number before which the exception is thrown. $e$ is the type of this exception.

\paragraph{Fault model} In this paper, we consider two fault models: 1) injecting only one exception, when the perturbation point is reached for the first time and 2) always injecting exceptions when the perturbation point is reached.

\paragraph{Perturbation search space} We define the ``perturbation search space'' as the Cartesian product of all possible perturbation points and all fault models with respect to a workload \cite{Danglot:correctness_attraction,Durieux:Exhaustive_Exploration_of_the_Failure-Oblivious_Computing_Search_Space}. The size of the search space is the number of workload executions required to have an exhaustive picture of the behavior under perturbation.

\paragraph{Exception handling method}
When an exception $e$ is handled in a method, this method is called the ``exception handling method'' for $e$. In a tuple $<$exception source, exception type, method$>$, the source refers to the location where the exception was thrown.
In this paper, we make a distinction between ``default exception handling methods'' and ``failure-oblivious method'':
the former refers to methods with manually written catch blocks while the latter refers to method with automatically instrumented catch blocks.

\paragraph{Acceptability oracle}
An acceptability oracle is a mechanism for determining whether an application's behaviour remains acceptable under perturbation.

In order to evaluate and improve an application's resilience, we use oracles to describe acceptable behaviour, hence we call them ``acceptability oracle''. In this paper, an acceptability oracle is a combination of generic oracles (like the absence of crash) and domain-specific ones. For example, in the context of a file downloading client, an acceptability oracle could be that 1) the client does not crash and exits normally 2) the client successfully downloads the file with a correct checksum.

\paragraph{Failure-oblivious method}
A method $fo$ is said to be failure-oblivious with respect to a perturbation point $p$ if and only if: 
1) when an exception is thrown from $p$, it is possible to catch and stop its propagation in $fo$ that is upper in the call stack;
2) the behaviour of the application verifies the acceptability oracle if the exception thrown from $p$ is caught in $fo$. This is noted $<m, l, e> \mapsto fo$ (here $\mapsto$ denotes ``thrown exception $e$ in method $m$ before location $l$ and caught at $fo$'').

\paragraph{Fault injection experiment} Given an application $a$ and a workload $w$, injecting an exception during the execution of $a$ under $w$ is called an ``experiment''.
\tripleagent is designed to conduct experiments in order to evaluate and improve an application's error-handling capability.

\emph{Example.} Let us assume an invocation chain across three methods: $m2 \rightarrow m1 \rightarrow m0$. Method $m0$ can throw an IOException before line number $l$, and the developers write a try-catch block to handle it in $m2$ (the method upper in the stack). 
Consequently, $m2$ is the default exception handling method for this exception.
If the exception is caught and silenced in method $m1$ and the application behaviour is still acceptable according to the oracle, method $m1$ is considered as a failure-oblivious method for $<m0, l, IOException>$.

\subsection{Goals of \tripleagent}\label{sec:goals}

\tripleagent aims at improving the exception-handling capabilities of Java applications. The main goals of \tripleagent are:
1) to give developers feedback about the effectiveness of their exception-handling design;
2) to automatically identify improvements of exception handling.

The former is about detecting the weakness points of the system under consideration and the latter is about finding new failure-oblivious methods that improve the application's resilience. 

\textbf{Input to \tripleagentbf:}
\tripleagent takes arbitrary software written in Java as input and a workload. No manual change is required from the developer. Neither source code nor test suite are required for improving an application's resilience.
\tripleagent also takes as input an acceptability oracle, which will be explained below in \autoref{sec:definitions}.

\textbf{Output for the developer:}
The output of \tripleagent is a report for developers. The report gives three pieces of information:
1) the perturbation points and their classification as defined next;
2) the verified failure-oblivious methods, i.e. the resilience improvements;
3) a log file which contains all the monitored information for the purpose of further analysis.

\tripleagent classifies the perturbation points into three categories as follows:

\begin{definition}{Fragile points:}
A fragile point is a statement in a method before which injecting one exception results in the application crashing or freezing.
\end{definition}

\begin{definition}{Sensitive points:}
A sensitive point is a statement in a method before which injecting one single exception does not influence the application in the workload under consideration. But, continuously injecting exceptions results in the application to crash or freeze.
\end{definition}

\begin{definition}{Immunized points:}
An immunized point is a statement in a method before which no matter how many exceptions are injected, the application still behaves acceptably.
\end{definition}

\subsection{Core Algorithm}\label{sec:algorithm}

\begin{algorithm}[tb]
\caption{Detection of Perturbation Points}
\label{algo:p-points-detection}
\begin{algorithmic}[1]
\REQUIRE ~~\\ 
An application $A$;\\
A repeatable workload for this application $W$;
\ENSURE ~~\\ 
$P$, a set of perturbation points;

\STATE Execute the application $A$ normally under $W$;
\FOR{each method $m$ loaded into the JVM}
\FOR{each checked exception $e$ thrown from $m$}
\FOR{each location $l \in m$}
    \STATE $P \leftarrow P \cup \{<m, l, e>\}$;
\ENDFOR
\ENDFOR
\ENDFOR
\RETURN $P$;
\end{algorithmic}
\end{algorithm}

\begin{algorithm}[tb]
\caption{Automated Classification of Perturbation Points}
\label{algo:p-points-classification}
\begin{algorithmic}[1]
\REQUIRE ~~\\ 
An application $A$;\\
A repeatable workload for this application $W$;\\
A set of perturbation points $P$;\\
An acceptability oracle $O$;
\ENSURE ~~\\ 
$F$ a set of fragile points points;\\
$S$ a set of sensitive points;\\
$I$ a set of immunized points;

\FOR{each point $<m, l, e> \in P$}
\STATE Execute the application under $W$;
\IF{$m$ is executed for the first time}
  \STATE Inject the exception $e$ before location $l$;
  \STATE $b1 \leftarrow$ application behaviour;
\ENDIF
\STATE Execute the application under $W$;
\STATE Always inject exceptions $e$ before location $l$;
\STATE $b2 \leftarrow$ application behaviour;
\IF{$b1$ not meet O and $b2$ not meet $O$}
  \STATE $F \leftarrow F \cup \{<m, l, e>\}$
\ELSIF{$b1$ meet $O$ and $b2$ not meet $O$}
  \STATE $S \leftarrow S \cup \{<m, l, e>\}$
\ELSIF{$b1$ meet $O$ and $b2$ meet $O$}
  \STATE $I \leftarrow I \cup \{<m, l, e>\}$
\ENDIF
\ENDFOR
\RETURN $F, S, I$;
\end{algorithmic}
\end{algorithm}

\begin{algorithm}[htb]
\caption{Detection of Failure-oblivious Methods}
\label{algo:fo-methods-detection}
\begin{algorithmic}[1]
\REQUIRE ~~\\ 
An application $A$;\\
A repeatable workload for this application $W$;\\
A set of perturbation points $P$;\\
An acceptability oracle $O$;
\ENSURE ~~\\ 
A set of failure-oblivious methods $R$;

\STATE // Find candidate failure-oblivious methods to be assessed
\STATE $Q \leftarrow \emptyset$ // Worklist for candidate methods;
\FOR{each point $<m, l, e> \in P$}
\STATE Execute the application under $W$;
\IF{ $m$ is executed for the first time}
  \STATE Inject the exception $e$ before location $l$;
  \FOR{each method $n$ in the call stack}
    \STATE $Q \leftarrow Q \cup \{<m, l, e> \mapsto n\}$;
  \ENDFOR
\ENDIF
\ENDFOR

\STATE
\STATE // Assess all candidate failure-oblivious methods
\FOR{each perturbation point $<m, l, e> \in P$}
\FOR{$<m, l, e> \mapsto n \in Q$}
\STATE Execute the application under $W$ twice as specified in~\autoref{algo:p-points-classification};
\STATE When $e$ is thrown, catch the exception $e$ in $n$;
\IF{the behaviour meets $O$}
  \STATE $R \leftarrow R \cup (<m, l, e>  \mapsto n)$;
\ENDIF
\ENDFOR
\ENDFOR

\RETURN $R$;
\end{algorithmic}
\end{algorithm}

The whole procedure of \tripleagent to identify failure-oblivious methods could be split into 3 steps:

1) Define acceptability oracles and detect perturbation points. \tripleagent executes the application normally, in order to monitor and record the application's normal behaviour. With this execution, the perturbation agent goes through all classes loaded into JVM and locates every  perturbation point, based on method signature, which is described in \autoref{algo:p-points-detection}.

2) Classify all the perturbation points into fragile, sensitive or immunized ones (as defined in \autoref{sec:goals}). For each perturbation point, \tripleagent conducts two experiments: only injecting one exception when the point is reached for the first time, and always injecting exceptions when the point is reached. Based on the observation of the application behaviour under perturbation, the perturbation point is classified using \autoref{algo:p-points-classification}.

3) Identify candidate failure-oblivious methods and evaluate each of them as described in \autoref{algo:fo-methods-detection}. \tripleagent detects candidate failure-oblivious methods with call stack analysis: every method in the stack before the default handling method is identified as a candidate failure-oblivious method. Then two fault injection experiments are conducted (only inject one exception, inject several exceptions). The difference is that all the thrown exceptions are caught in a catch block instrumented in candidate failure-oblivious methods. By analyzing the behavior once the exception is caught in this catch block, \tripleagent confirms whether the method under evaluation is indeed failure-oblivious or not.

\subsection{Architecture of \tripleagent}\label{sec:components}

\begin{figure*}[tb]
\centering
\includegraphics[width=17.5cm]{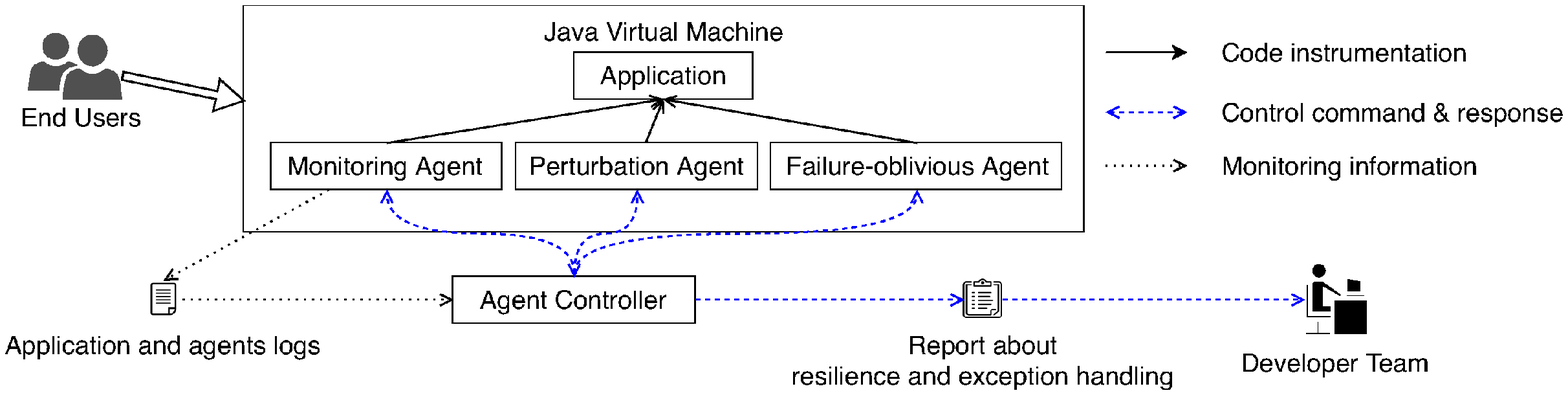}
\caption{The components of \tripleagent}\label{fig:components}
\vspace{-0.3cm}
\end{figure*}

\autoref{fig:components} shows the general architecture of \tripleagent. 
\tripleagent considers a Java application in a JVM, such as a backend web application or a Java micro-service.

When an application is loaded into the JVM, \tripleagent attaches to it three different agents: a monitoring agent, a perturbation agent and a failure-oblivious agent. The monitoring agent is responsible for collecting the information needed by \tripleagent to evaluate the system's resilience capabilities. The perturbation agent injects exceptions into the application in order to trigger its error-handling logic. The failure-oblivious agent tries to improve the application's resilience by catching and silencing exceptions before they are handled by default exception handling methods.

All the agents are controlled by a controller which makes two kinds of decisions:
1) given an application under some specific workload, which perturbation point should be activated,
2) whether the point's corresponding failure-oblivious method should be switched on.
Finally, the controller generates a report for the developer based on data gathered from a series of fault injection experiments.

\subsubsection{Monitoring agent}
In order to study the influence of perturbations and evaluate all possible failure-oblivious methods in a software system, it is necessary to collect different kinds of monitoring information. For this, we propose to use a monitoring agent that is attached to the runtime process.

Our monitoring agent works as follows.
For each method in the code loaded in the JVM, the agent collects static and dynamic information.

The static information is:
1) its position in the code,
2) whether it declares checked exceptions to be thrown.

The collected dynamic information is:
3) the number of method executions over an fault injection experiment,
4) each time an exception is caught, the agent collects the stack information, including the stack distance between the method raising the exception and the method catching it. This includes both exceptions caught in default exception handlers and in failure-oblivious methods (as defined in \autoref{sec:definitions}).

The \tripleagent monitoring agent also collects the following information:
\begin{itemize}
\item The set of classes that have been loaded into the JVM.
\item Whether the application has exited normally or crashed due to an unhandled exception.
\end{itemize}

\subsubsection{Perturbation agent}
The perturbation agent injects specific perturbations at a specific point in time. The perturbation commands come from the agent controller.

The perturbation agent detects every method with a throws keyword and attaches itself into this method by rewriting the bytecode. In order to explore the entire perturbation search space, the agent injects different perturbation points before each statement in the method. In this way, the agent is able to throw such an exception anywhere in the method and compare the difference.

\begin{lstlisting}[caption={The Perturbation Strategy in \tripleagent},label=lst:perturbationAgent]
// a perturbation point in Class1
void exampleMethod() throws ExceptionA, ExceptionB {
  // code injected with code transformation
  PAgent.throwExceptionPerturbation(key1);
  PAgent.throwExceptionPerturbation(key2);
  // a statement
  PAgent.throwExceptionPerturbation(key3);
  PAgent.throwExceptionPerturbation(key4);
}
\end{lstlisting}

\autoref{lst:perturbationAgent} gives an example of how this perturbation agent works. When a method like exampleMethod() throws multiple exceptions, corresponding perturbation points are automatically injected with code transformation. The perturbation agent controls every perturbation point separately. When a specific point is activated, it throws an exception at the beginning of the method.

\subsubsection{Failure-oblivious agent}\label{sec:fo-agent}
The failure-oblivious agent instruments the code with try-catch blocks during a fault injection experiment.
For reasoning about resilience with respect to uncaught exceptions, the failure-oblivious agent injects a try-catch wrapper in all methods. Basically, the whole method body is wrapped with a try-catch block which handles all types of exceptions (\texttt{catch Exception} in Java).
By default, the catch block simply throws again the exception which makes it semantically equivalent to the original code.
When the failure-oblivious method is activated, the injected catch block silences the exception and prevents it from propagating (note that the exception may come from this method or from other methods transitively called from this method).

When an exception is caught by the injected catch block, there are three possible outcomes:
1) the application runs normally;
2) the application runs in a gracefully degraded mode;
3) the application crashes.

\begin{lstlisting}[caption={Automated Code Instrumentation for Identifying Failure-oblivious Methods in \tripleagent},label=lst:foAgent]
// a candidate failure-oblivious method in Class2
void callExampleMethod() throws ExceptionA, ExceptionB {
  try {
    new Class1().exampleMethod();
  } catch (Exception a) {
    if (!FOAgent.modeIsOn(key2)) {
      throw a;
    } else { // nothing, the exception is silenced }
  }
}
\end{lstlisting}
\vspace{0.3cm}

\autoref{lst:foAgent} illustrates how this is done. In method $callExampleMethod$,  $exampleMethod$ is invoked. The failure-oblivious agent detects it as a possible failure-oblivious method. So the whole method body of $callExampleMethod$ is wrapped with a try catch block. When the agent controller activates this failure-oblivious method, it silences all exceptions coming from $exampleMethod$. Otherwise it throws the caught exception so that it is propagated as usual.

\subsubsection{Agents controller}\label{sec:agent-controller}
The agent controller is responsible for conducting a series of experiments (see \autoref{sec:definitions}). It controls every agent and gathers all the information to analyze the system resilience. Additionally, the controller is configurable. For example, developers can define a filter to focus on resilience improvement for a specific package.

\subsection{Implementation}
There are different kinds of agents in the JVM.
The monitoring agent is implemented on top of the JVM Tool Interface (JVMTI)~\footnote{See \url{https://docs.oracle.com/javase/8/docs/platform/jvmti/jvmti.html}}. The perturbation agent and failure-oblivious agent are implemented as JVM agents, using the ASM library for binary code transformation~\footnote{See \url{http://asm.ow2.org}}. The agents controller is a standalone service, it communicates with the JVM and the agents through local files.


For sake of open-science, the code is made publicly available at \url{http://bit.ly/tripleagent-repo}.


\section{Evaluation}\label{sec:evaluation}

For evaluating this contribution, we apply a case-based evaluation methodology: this methodology consists of an in-depth analysis of relevant cases selected in a principled way~\cite{flyvbjerg2006five}. In our research domain, it has been shown appropriate in Rinard et al's original paper on failure-oblivious computing \cite{rinard2004enhancing}. 

We select two case studies according to the following three criteria:
1) the case should be a real-world application (i.e., not a toy example)
2) it should be medium-sized in order to be appropriate for the computing power available in the laboratory
3) it is possible to define a production-like workload.
Those criteria yield two cases: TTorrent and HedWig. TTorrent is a file transferring tool which implements the BitTorrent protocol. HedWig is an email server for the IMAP, SMTP and POP3 protocols. They are also exemplary of applications with high resilience requirements: an email server must not crash, a file download on the dynamic internet must succeed regardless of unexpected network events, peer failures, or local machine issues.

The analysis of \tripleagent requires several executions. For each perturbation point, its classification requires $2$ executions under the workload (as discussed in~\autoref{algo:p-points-classification}).
For each candidate failure-oblivious method, its evaluation also needs $2$ executions  under the workload  (see \autoref{algo:fo-methods-detection}). 
For both cases, an execution takes no more than $1$ minute in our testing environment. In total, the cost of the  experiments presented in this section is upper-bounded by $2 \times 1 \times (1046 + 2844 + 372 + 722) = 9968$ minutes. Note that some experiments lead applications to a crash. It actually takes \tripleagent around $3$ days to finish all the experiments.
\subsection{Evaluation on TTorrent}

\subsubsection{Experiment Protocol}

We apply \autoref{sec:algorithm} to TTorrent, version \href{https://github.com/mpetazzoni/ttorrent/releases/tag/ttorrent-2.0}{2.0}.

The workload $W$ for TTorrent consists of downloading a large file (\href{https://cdimage.debian.org/debian-cd/current/amd64/bt-cd/}{debian-9.9.0-amd64-netinst.iso}, a Debian distribution installer of 292.0MB).
Since BitTorrent is a peer-to-peer protocol, this workload involves other machines on the internet which serve (aka "seed") the file. To that extent, the workload is a production one.
We perform a series of fault injection experiments, as described in \autoref{sec:definitions}.

For TTorrent, we consider the following definition of acceptable behaviour to evaluate candidate failure-oblivious methods:
the behaviour is considered acceptable if an end-user can successfully download a file with a correct checksum.

\subsubsection{Experimental Results}\label{sec:results-ttorrent}

Per \autoref{sec:algorithm}, the first step of \tripleagent is to execute TTorrent normally and to monitor all possible perturbation points.
It detects $1046$ points in total within the package \texttt{com/turn/ttorrent}.

Then, \tripleagent performs two series of experiments:
1) it injects one exception per perturbation point and compares the behaviour between these experiments and the normal execution,
2) it always injects exceptions when a perturbation point is reached and also compares the behaviour against the reference one.

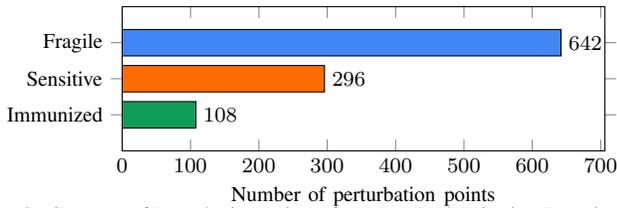
\begin{figure}[tb]
\centering
\begin{tikzpicture}
  \footnotesize
  \begin{axis}[
        xbar, xmin=0,
        width=8cm, height=3.5cm, enlarge y limits=0.5,
        xlabel={Number of perturbation points},
        ytick={1, 2, 3},
        yticklabels={Immunized, Sensitive, Fragile},
        nodes near coords, nodes near coords align={horizontal},
        every axis plot/.append style={
          bar shift=0pt
        }
    ]
    \addplot[fill=fragile-points] coordinates {(642,3)};
    \addplot[fill=sensitive-points] coordinates {(296,2)};
    \addplot[fill=immunized-points] coordinates {(108,1)};
  \end{axis}
\end{tikzpicture}
\vspace{-0.5cm}
\caption{Category of Perturbation Points after $2092$ Fault Injection Experiments on TTorrent}\label{fig:result-classification}
\end{figure}

As a result, all perturbation points get classified in the 3 categories defined in \autoref{sec:theDesignOfTripleAgent}. In total, \tripleagent identifies  $642$ fragile points, $296$ sensitive points and $108$ immunized points. \autoref{fig:result-classification} shows the distribution of these perturbation points, which are used as a base line for the following experiments.

The next step of \tripleagent's main algorithm is to compute and assess the possible failure-oblivious methods.
As explained in~\autoref{algo:fo-methods-detection}, for a perturbation point, a set of failure-oblivious methods is identified. In our experiment, \tripleagent detects $2844$ possible failure-oblivious methods, summed over all the perturbation points. The minimum, median and maximum number of candidate failure-oblivious methods per perturbation point is respectively 0, 2, 10.

Then, $2844 \times 2 = 5688$ executions are made to assess the failure-obliviousness of the candidate points (one per injection mode). Hopefully, the added catch blocks inserted by \tripleagent will increase the number of immunized points. 

\begin{figure}[tb]
\centering
\includegraphics[width=8.5cm]{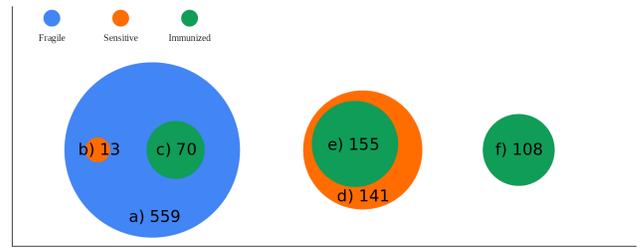}
\leftline{\scriptsize{a) Fragile stays fragile, b) Fragile to sensitive, c) Fragile to immunized}}
\leftline{\scriptsize{d) Sensitive stays sensitive, e) Sensitive to immunized, f) Immunized stays immunized}}
\vspace{-0.3cm}
\caption{Resilience Improvement on TTorrent: Fragile points, sensitive points and immunized points are respectively shown in blue, orange and green. The area of bubbles are corresponding to the numbers of perturbation points under consideration.}\label{fig:resilience-improvement-ttorrent}
\vspace{-0.5cm}
\end{figure}

Let us now consider \autoref{fig:resilience-improvement-ttorrent}.
The fragile, sensitive and immunized perturbation points are respectively shown in blue, orange and green. The area of bubbles corresponds to the numbers of perturbation points under consideration. For example, the bubble \texttt{e} represents the $155$ sensitive points transformed into immunized ones with failure-oblivious computing.
Overall, \tripleagent successfully transforms $13$ fragile points into sensitive ones, $70$ fragile points into immunized ones and $155$ sensitive points into immunized ones. The original $108$ immunized points remain immunized. This means that resilience of the TTorrent has been automatically improved.

\autoref{tab:failure-oblivious-improvement} presents a sample of perturbation points. Every row describes 
1) a perturbation point (the class name, method name and its line number), the thrown exception type, and the corresponding default exception handler written by developers;
2) the failure-oblivious improvement (failure-oblivious method and concrete change of the perturbation point's category).
For example, row 1 and row 2 show that \tripleagent verifies failure-oblivious methods which improve the original fragile perturbation points into sensitive ones. Row 7 and row 8 also describe the case that \tripleagent is able to detect multiple failure-oblivious methods for the same perturbation point. For original immunized perturbation points, alternative failure-oblivious methods which provide the same resilience are verified as well, which is shown in the last two rows.

\begin{table*}[tb]
\centering
\scriptsize
\begin{tabular}{@{\makebox[3em][r]{\stepcounter{rowcount}\therowcount\hspace*{\tabcolsep}}}lllll}
\toprule
{\bf Perturbation Point}& {\bf Exception Type}& {\bf Default Handling Method} & {\bf Failure-oblivious Method} & {\bf Improvement}\\
\midrule
BEValue/getNumber@122& InvalidBEnException& ClientMain/main& BEValue/getLong& fragile - sensitive\\
HTTPTrackerC/encodeAnnoToURL@187& AnnoException& TrackerClient/annoAllInterfaces& HTTPTrackerC/announce& fragile - sensitive\\
CommuManager/addTorrent@229& IOException& ClientMain/main& CommuManager/addTorrent& fragile - immunized\\
TorrentParser/getStringOrNull@121& InvalidBEnException& ClientMain/main& TorrentParser/getStringOrNull& fragile - immunized\\
HTTPTrackerC/sendAnnounce@235& ConnectException& HTTPTrackerClient/announce& HTTPTrackerC/sendAnnounce& fragile - immunized\\
SharedTorrent/init@226& InterruptedException& SharedTorrent/initIfNecessary& SharedTorrent/init& sensitive - immunized\\
SharedTor/handlePieceCompleted@671& IOException& SharingPeer/handleMessage& SharedTor/handlePieceCompleted& sensitive - immunized\\
SharedTor/handlePieceCompleted@671& IOException& SharingPeer/handleMessage& SharingPeer/firePieceCompleted& sensitive - immunized\\
WorkingReceiver/processAndGetNext@64 & IOException& ConnWorker/processSelectedKeys& ReadableKeyProcessor/process& sensitive - immunized\\
SharingPeer/send@352& IllegalStateException& CommuManager/validatePieceAsync& SharingPeer/send& alternative resilient method\\
PeerMessage/parse@176& ParseException& ConnWorker/processSelectedKeys& PeerMessage/parse& alternative resilient method\\
\bottomrule
\end{tabular}
\caption{Sample of Perturbation Points and the Corresponding Failure-Oblivious Methods in TTorrent}\label{tab:failure-oblivious-improvement}
\vspace{-0.5cm}
\end{table*}

\subsubsection{Case Studies}

In the following, we detail 3 case studies where the resilience is improved.


\begin{lstlisting}[caption=IllegalStateException in SharingPeer/send,label=lst:ttorrentCase1]
void send(PeerMsg m) throws IllegalStateException {
  if (this.isConnected()) {
    ...
  } else {
    // perturbation point here
    unbind(true);
  }
  ...
}
\end{lstlisting}

\paragraph{Failure-oblivious Method as Alternative to Normal Resilience}
First, \autoref{lst:ttorrentCase1} shows a failure-oblivious method with respect to exception \texttt{IllegalStateException}. This method is executed only 1 time during normal download of the file. Under perturbation, \tripleagent identifies that if one single exception is thrown from this method, the application is still able to download the file correctly. By analyzing the stack, \tripleagent detects another two methods as candidate failure-oblivious methods: \texttt{SharingPeer/send} and \texttt{SharingPeer/notInteresting}.

By activating a failure-oblivious try-catch block in these two methods (i.e. the method body is wrapped with a try-catch block which blocks the exception), TTorrent still succeeds in downloading the file. It means that \tripleagent successfully detects 2 alternative methods in the stack that provide the same resilience as the original manually-written catch block.

\begin{lstlisting}[caption=IOException in WorkingReceiver/processAndGetNext,label=lst:ttorrentCase2]
DataProcessor processAndGetNext(ByteChannel sc) throws IOException {
  ...
  // perturbation point here
  if (this.pstrLength > MAX_MESSAGE_SIZE) {
    return new ShutdownAndRemovePeerProcessor(...).processAndGetNext(socketChannel);
  }
}
\end{lstlisting}

\paragraph{Improving Resilience under a High Number of Exceptions}
\autoref{lst:ttorrentCase2} shows method \texttt{processAndGetNext} in class \texttt{WorkingReceiver}. This method is executed $34304$ times during the reference execution. If \tripleagent injects one single exception in this method when it is called for the first time, the application still downloads the file correctly. However, when the perturbation agent keeping injecting exceptions when downloading the file, the application gets stalled. 

After analyzing the call stack, \tripleagent detects $4$ candidate failure-oblivious methods, namely \texttt{WorkingReceiver/processAndGetNext}, \texttt{OutgoingConnectionListener/onNewDataAvail}, \texttt{ReadableKeyProcessor/process} and \texttt{ConnectionWorker/processSelectedKey}.
After two fault injection experiments per candidate failure-oblivious method, \tripleagent observes that the last three methods are failure-oblivious, the application downloads the file successfully, no matter how many exceptions are thrown in \texttt{processAndGetNext}. 

In this case, \tripleagent succeeds in detecting $3$ failure-oblivious methods that provide better resilience compared to the normal error-handling code written by the developer.

\begin{lstlisting}[caption=InvalidBEncodingException in TorrentParser/getStringOrNull,label=lst:ttorrentCase3]
String getStringOrNull(Map<...> m, String k) throws InvalidBEncodingException {
  // perturbation point here
  BEValue v = dictionaryMetadata.get(key);
  if (v == null) return null;
  return v.getString();
}
\end{lstlisting}

\paragraph{Improving Resilience from Crashing to Resilient}
Let now us consider \autoref{lst:ttorrentCase3}. With a fault injection experiment in method \texttt{getStringOrNull} before line 3, \tripleagent identifies that an exception \texttt{InvalidBEncodingException} thrown at this location crashes the whole process. Hence, the perturbation point is a fragile one. After analyzing the stack information, \texttt{ClientMain/main} is the default handling method. There are 7 methods including \texttt{getStringOrNull} itself before this default handling method, which are all considered as candidate failure-oblivious methods by \tripleagent.

Then, \tripleagent performs 2 fault injection experiments for each method according to \autoref{algo:fo-methods-detection}, that is $2 \times 7 = 14$ experiments in total. The first experiment assesses whether the candidate failure-oblivious methods could handle only one injected exception. The second assesses whether they could handle as many as injected exceptions.
Indeed, \tripleagent observes that when a catch block is automatically injected in \texttt{getStringOrNull}, the application does not crash anymore, and even better, the resulting behaviour is correct (the file is correctly downloaded, its content is the expected one, bit-per-bit).
In this case, \tripleagent has automatically transformed a crashing exception into acceptable behaviour.

\begin{mdframed}[style=mpdframe,frametitle=Insights from the TTorrent experiment]
Under a realistic workload of downloading a 200MB+ file from the internet, \tripleagent performs $7780$ experiments to evaluate $1046$ perturbation points spread over $6.5$kLOC. 
\tripleagent identifies $642$ fragile points, $296$ sensitive points and $108$ immunized points.
After analyzing all $2844$ candidate failure-oblivious methods, \tripleagent confirms that there are $238$ failure-oblivious methods in the application.
This shows that it is feasible to automatically improve resilience by combining perturbation injection and failure-obliviousness analysis.
\end{mdframed}

\subsection{Evaluation on HedWig}
\subsubsection{Experiment Protocol}
HedWig is an email server written in Java, a typical server side application. The main process of HedWig is a perpetual loop, which creates sub-threads to handle different user requests. HedWig relies on a MySql database to store email metadata and saves the email contents as files on the disk. In this experiment, we consider the latest version of HedWig (\href{https://sourceforge.net/projects/hwmail/files/0.7/}{v0.7}).

The considered workload is as follows.
First, HedWig is deployed on a server. Then \tripleagent sends a specific email with a unique content to a testing email address using the SMTP protocol. Finally, \tripleagent logs in with the corresponding account and fetches the same email to do the comparison. The acceptable behaviour is that \tripleagent both successfully sends and fetches the email, and that the content of this email after final fetching is totally correct.

The experiments are performed sequentially. We note that some of the perturbation experiments crash the email server. In this case, the server is automatically restarted. 
Some perturbation experiments put it in a corrupted state: to detect this, \tripleagent adds a checking point after each experiment. All the perturbation agents are switched off and an email is sent and fetched as usual. If the server works correctly the next experiment goes on, otherwise \tripleagent runs a restart script to bring the server back to normal state.

\subsubsection{Experimental Results}\label{sec:results-hedwig}

\begin{figure}[tb]
\centering
\begin{tikzpicture}
  \footnotesize
  \begin{axis}[
        xbar, xmin=0,
        width=8cm, height=3.5cm, enlarge y limits=0.5,
        xlabel={Number of perturbation points},
        ytick={1, 2, 3},
        yticklabels={Immunized, Sensitive, Fragile},
        nodes near coords, nodes near coords align={horizontal},
        every axis plot/.append style={
          bar shift=0pt
        }
    ]
    \addplot[fill=fragile-points] coordinates {(264,3)};
    \addplot[fill=sensitive-points] coordinates {(14,2)};
    \addplot[fill=immunized-points] coordinates {(94,1)};
  \end{axis}
\end{tikzpicture}
\caption{Category of Perturbation Points under 744 Fault Injection Experiments on HedWig}\label{fig:result-hedwig-classification}
\end{figure}
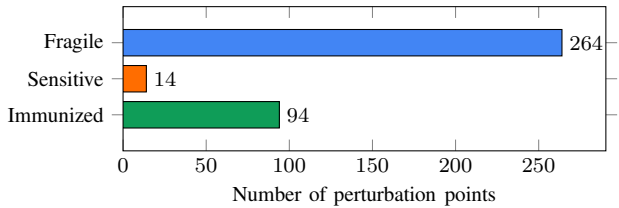

Within the package \texttt{com/hs/mail} \tripleagent detects $372$ perturbation points. Each perturbation point is evaluated by two fault injection experiments: 1) only one exception is injected during the email sending and fetching process, when the point is reached for the first time and 2) exceptions are always injected when the point is reached. Based on these $744$ experiments \tripleagent classifies all the perturbation points using the classification algorithm described in \autoref{algo:p-points-classification}. Overall, \tripleagent finds in Hedwig $264$ fragile points, $14$ sensitive points and $94$ immunized points, which are shown in~\autoref{fig:result-hedwig-classification}

The next step for \tripleagent is to identify the candidate failure-oblivious methods. By summing over all perturbation points, \tripleagent detects $722$ candidate methods. The minimum, median, maximum number of candidate failure-oblivious methods per perturbation point is respectively 0, 2, 10.

\begin{figure}[tb]
\centering
\includegraphics[width=8.5cm]{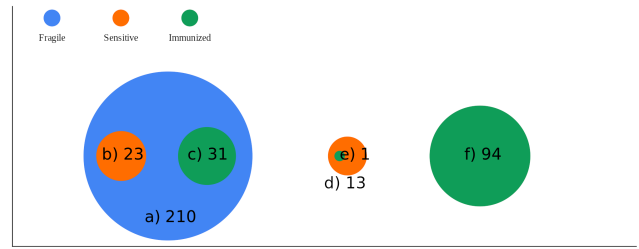}
\leftline{\scriptsize{a) Fragile stays fragile, b) Fragile to sensitive, c) Fragile to immunized}}
\leftline{\scriptsize{d) Sensitive stays sensitive, e) Sensitive to immunized, f) Immunized stays immunized}}
\vspace{-0.5cm}
\caption{Resilience Improvement on HedWig}\label{fig:resilience-improvement-hedwig}
\vspace{-0.5cm}
\end{figure}

Similar to classifying perturbation points, each candidate failure-oblivious method also needs two fault injection experiments to be evaluated. Finally, $1444$ executions are made to evaluate all the candidate failure-oblivious methods based on \autoref{algo:fo-methods-detection}. By silencing exceptions in the candidate failure-oblivious methods, \tripleagent shows that $23$ fragile perturbation points can be improved into sensitive ones. $31$ fragile points are transformed to immunized ones. It upgrades $1$ sensitive perturbation point to an immunized one as well. All those improvements are shown in \autoref{fig:resilience-improvement-hedwig}.

\autoref{tab:failure-oblivious-improvement-hedwig} shows a sample of interesting perturbation points. It shows different levels of automatic resilience improvement. Similar to \autoref{tab:failure-oblivious-improvement}, every row describes one perturbation point and one of its corresponding failure-oblivious methods. For example, the first row gives details about perturbation point \texttt{queryForLong} in class \texttt{AbstractDao}, line number $100$. When a \texttt{DataAccessException} is thrown from this point, by default it is handled by a try-catch block written by developers in \texttt{TransactionTemplate/execute}. But this catch block does not prevent the exception from failing user requests. If the same exception is caught earlier in the stack, in \texttt{AnsiMessageDao/getHeaderNameID}, the server is able to bear at least one exception. Note that it is possible to have multiple failure-oblivious methods for the same perturbation point. Such as row 3 and row 4, row 8 and row 9 in the table.

\begin{table*}[tb]
\centering
\scriptsize
\setcounter{rowcount}{-1}
\begin{tabular}{@{\makebox[3em][r]{\stepcounter{rowcount}\therowcount\hspace*{\tabcolsep}}}lllll}
\toprule
{\bf Perturbation Point}& {\bf Exception Type}& {\bf Default Handling Method} & {\bf Failure-oblivious Method} & {\bf Category}\\
\midrule
AbstractDao/queryForLong@100& DataAccessException& TransactionTemplate/execute& AnsiMessageDao/getHeaderNameID& fragile - sensitive\\
ImapServerHandler/handleUpstream@56& Exception& DCPipeline/sendUpstream& ImapServerHandler/handleUpstream& fragile - sensitive\\
CountingInputStream/read@21& IOException& BodySBuilder/build& BodySBuilder/simplePartDescriptor& fragile - immunized\\
CountingInputStream/read@21& IOException& BodySBuilder/build& BodySBuilder/createDescriptor& fragile - immunized\\
MessageHeader/parse@118& IOException& MessageHeader/<init>& MessageHeader/parse& fragile - immunized\\
FlagUtils/getFlags@57& SQLException& JdbcT/doInPreparedStatement& FlagUtils/getFlags& fragile - immunized\\
PartContentBuilder/build@63& IOException& FetchRespBuilder/bodyContent& PartContentBuilder/build& fragile - immunized\\
MailMessage/save@88& IOException& ToRepository/service& MailMessage/save& sensitive - immunized\\
MailMessage/save@88& IOException& ToRepository/service& ToRepository/saveMessage& sensitive - immunized\\
AliasingForwarding/service@70& MessagingException& LocalDelivery/service& AliasingForwarding/service& alternative resilient method\\
ToRepository/deliver@118& IOException& ToRepository/service& ToRepository/deliver& alternative resilient method\\
\bottomrule
\end{tabular}
\caption{Sample of Perturbation Points and the Corresponding Failure-Oblivious Methods in HedWig}\label{tab:failure-oblivious-improvement-hedwig}
\vspace{-0.8cm}
\end{table*}

\subsubsection{Case Studies}

We now discuss two interesting case studies.

\begin{lstlisting}[caption=SQLException in AnsiMailboxDao\$1/mapRow,label=lst:hedwigCase1]
mailboxRowMapper = new RowMapper<Mailbox>() {
  public Mailbox mapRow(ResultSet r, int n) throws SQLException {
    // perturbation point here
    Mailbox mb = new Mailbox();
    mb.setMailboxID(r.getLong("mailboxid"));
    ...
    return mb;
  }
};
\end{lstlisting}

\paragraph{A Failed Failure-oblivious Experiment}
\autoref{lst:hedwigCase1} shows a perturbation point found by \tripleagent in Class \texttt{AnsiMailboxDao}, line 3. First, \tripleagent detects that when one \texttt{SQLException} is thrown from this location, the application fails to receive and send the test email. Hence, the perturbation point is a fragile one. By analyzing the call stack, method \texttt{mapRow} is considered as a candidate failure-oblivious method. 

\tripleagent automatically wraps the method with a try-catch block. This specific failure-oblivious operation results in inserting an incorrect record into the database, which influences the upcoming experiments. Thanks to running the checkpoint procedure described above, \tripleagent detects this problem, restarts the server, definitely labels this method as non failure-oblivious and excludes this perturbation point for later experiments.

\begin{lstlisting}[caption=IOException in CountingInputStream/read,label=lst:hedwigCase2]
public int read() throws IOException {
  // perturbation point here
  int next = in.read();
  ...
  return next;
}
\end{lstlisting}

\paragraph{A Perturbation Point with Multiple Failure-oblivious Methods}
In \autoref{lst:hedwigCase2}, line 2 is a fragile perturbation point in Class \texttt{CountingInputStream}. If an \texttt{IOException} is thrown from this location, the user is not able to fetch any emails. By default the exception is handled by \texttt{BodyStructureBuilder/build(Date d, Long l)}, 5 methods upper in the stack. It means that methods before the default exception handler are all candidate failure-oblivious methods. \tripleagent evaluates them one by one and verifies 3 out of them. In the call stack, if the exception is silenced in \texttt{BodyStructureBuilder/simplePartDescriptor}, \texttt{BodyStructureBuilder/createDescriptor} or \texttt{BodyStructureBuilder/build(InputStream i)}, the server works properly no matter how many exceptions are thrown. This is a strong improvement to the resilience.

\begin{mdframed}[style=mpdframe,frametitle=Insights from the HedWig experiment]
Under a production-like email task, \tripleagent performs $2188$ experiments to evaluate $372$ perturbation points spread over $13.8$ kLOC. $261$ fragile points, $68$ sensitive points and $43$ immunized points are identified in the original code. 
\tripleagent assesses that $60$ out of $722$ methods can be transformed into failure-oblivious methods.
This further confirms that \tripleagent can improve  the resilience of a server application in an automated manner.
\end{mdframed}

\subsection{Overhead of \tripleagent}\label{overhead}

The overhead of \tripleagent varies a lot among different perturbation points, failure-oblivious methods. Considering that the ultimate goal of \tripleagent is to automatically improve resilience, we manually evaluate the overhead of failure-oblivious experiments. The overhead caused by \tripleagent is evaluated in 3 aspects:
1) at the application level, the execution time is compared. In TTorrent this metric means the downloading time. In HedWig experiments this means the time spent on sending and receiving the email.
2) at the operating system level, the CPU and memory usage, peak thread count are taken into consideration.
3) at the binary code level, the code bloat due to instrumentation is evaluated.
For statistical purposes, we conduct the same measurement 30 times and calculate the average~\cite{Arcuri:A_practical_guide_for_using_statistical_tests_to_assess_randomized_algorithms_in_software_engineering}.

For TTorrent experiments, \autoref{tab:overheadOfTTorrent} records the overhead of verifying failure-oblivious method \texttt{HTTPTrackerClient/sendAnnounce} in \autoref{tab:failure-oblivious-improvement}, row 5. For HedWig experiments, failure-oblivious method \texttt{BodySBuilder/simplePartDescriptor} in \autoref{tab:failure-oblivious-improvement-hedwig}, row 3 is taken as an example. The overhead of execution time, CPU time, memory usage, peak thread count, relevant class files size are respectively 0\%, 6.0\%, 0\%, 5.4\%, 3.0\%.


\begin{table}[!tb]
\centering
\caption{The Overhead of An Experiment on TTorrent}\label{tab:overheadOfTTorrent}
\scriptsize
\begin{tabular}{lrrr}
\toprule
Evaluation Aspects& Original Version& Instrumented Version& Variation\\
\midrule
Downloading time& 20.4s& 21.1s& 3.5\%\\
CPU time& 15.0s& 18.3& 22.2\%\\
Memory usage& 47M& 49M& 4.3\%\\
Peak thread count& 30& 32& 6.7\%\\
Relevant class files size& 16.7KB& 16.8KB& 0.6\%\\
\bottomrule
\end{tabular}
\vspace{-0.5cm}
\end{table}

The reason why \tripleagent has such a low overhead is that the instrumentation is small. The perturbation agent and failure-oblivious agent only instrumented one or two class files. Meanwhile, the monitoring agent does not cause high overhead thanks to the JVMTI framework. By evaluating the overhead, developers are more confident about the resilience improvement suggested by \tripleagent.

\section{Discussion}
\emph{Fault model.} Currently \tripleagent considers two fault models. In both models, exceptions are injected at a single location. But there also exists common mode failures which involve a series of different exceptions. An exception could also be mixed with data errors. Devising and implementing fault models that stimulate common mode failures or data errors is an interesting direction for future work.

\emph{Workload impact.} A threat to the validity of our experiments comes from the workload. \tripleagent takes a production-like workload to exercise the application code. When \tripleagent identifies a failure-oblivious method, it guarantees that it works under the tested workload.
But it may break some behavior with a more comprehensive workload.
Overall, the more diverse the workload is, the more confidence \tripleagent has in the found failure-oblivious methods.

\emph{Scalability.} During our experiments, the deepest call stack considered was composed of 39 methods. \tripleagent has not been tested with larger applications with deeper stacks.
As such, the full scalability of \tripleagent is not yet verified. 
We note that the number of candidate failure-oblivious methods to assess is linear in the depth of the stack, which means that, in theory, \tripleagent would be scalable.

\section{Related work}\label{sec:relatedWork}
Now we discuss the related work along three aspects.

\subsubsection{Fault injection}
Fault injection is a widely-researched topic in software dependability. In the 1990s, the research was mostly about hardware implemented fault injection tools. For example, Madeira et al.~\cite{Madeira:RIFLE} invented RIFLE, a pin-level fault injector to generate processor errors. Next, more software-based fault injection tools were invented. 
Kanawati et al.~\cite{Ghani:FERRARI} proposed FERRARI, a tool for the emulation of hardware faults and control flow errors. Han et al.~\cite{Seungjae:DOCTOR} designed DOCTOR, a tool for injecting hardware failures and network communication failures.
Wei et al.~\cite{Wei:Quantifying_the_Accuracy_of_High-Level_Fault_Injection_Techniques_for_Hardware_Faults:DSN} built a software-based hardware faults injector called LLFI, and quantitatively compared the accuracy of fault-injection with assembly code level injector PINFI.
Lee et al.~\cite{Hyosoon:SFIDA_a_software_implemented_fault_injection_tool_for_distributed_dependable_applications} presented SFIDA, a tool to test the dependability of distributed applications on the Linux platform.
Kao et al.\cite{Kao:A_fault_injection_and_monitoring_environment_for_tracing_the_UNIX_system_behavior_under_faults} invented ``FINE'', a fault injection and monitoring tool to inject both hardware-induced software errors and software faults. Kouwe and Tanenbaum\cite{Kouwe:HSFI_Accurate_Fault_Injection_Scalable_to_Large_Code_Bases:DSN} presented HSFI, a fault injection tool that injects faults with context information from  source code and applies fault injection decisions efficiently on the binary.

Fu et al.\cite{Fu:Compiler-directed_program-fault_coverage_for_highly_available_Java_internet_services} presented an approach to measure the coverage of recovery code with respect to operating system and I/O hardware faults. The common idea with \tripleagent is to inject exceptions to trigger error handling code. Yet, our and their goal are notably different. Fu el al. use fault injection to increase recovery code coverage. \tripleagent combines fault injection with failure-oblivious computing to improve resilience.

The novelty of \tripleagent is that it is designed to inject application-level exceptions (and not hardware faults) in Java applications. \tripleagent gives developers concrete insights at the source code level about their exception-handling implementation.

\subsubsection{Self-healing software}
Self-healing software follows the idea that it is possible to automatically make software recover from failures. Different techniques have been applied to achieve this goal, such as automatic reboot, checkpoint-restart, and failure-oblivious transformation.

Reboot techniques~\cite{Candea:Crash-Only_Software,Huang:Software_Rejuvenation_Analysis_Module_and_Applications,Sullivan:Software_Defects_and_their_Impact_on_System_Availability} require the system to be able to restart, which may bring some down-time.
Checkpoint-restart techniques significantly reduce the recovery time by saving and reloading runtime states saved at checkpoints. Qin et al.~\cite{Qin:Rx_Treating_bugs_as_allergies_a_safe_method_to_survive_software_failures} invented Rx, which enables the program to rollback to a recent checkpoint upon a software failure, and then to re-execute in a modified environment.
Sidiroglou et al.~\cite{Sidiroglou:ASSURE} proposed ASSURE, a system that introduces rescue points to recover from unknown faults.

Regarding failure-oblivious computing,
Rinard et al.~\cite{rinard2004enhancing} invented a safe compiler for C to enable servers to execute despite memory errors.
Perkins et al.~\cite{Perkins:Automatically_patching_errors_in_deployed_software} proposed ClearView, a system for automatically patching errors in deployed software. It observes values of registers and memory locations and tries to detect violations of invariants at this level.
Rigger et al.~\cite{Rigger:Introspection_for_C_and_its_Applications_to_Library_Robustness} presented an approach that allows C programmers to perform explicit sanity checks and to react according to invalid arguments or states. They also designed a C dialect called Lenient C~\cite{Rigger:Lenient_Execution_of_C_on_a_Java_Virtual_Machine} that checks undefined behaviours in the C standard including memory management, pointer operations and arithmetic operations.

None of these tools combine fault injection and failure-oblivious computing together as we do in \tripleagent. They do not actively inject failures into the system, nor do they conduct application-level analysis to detect valuable failure-oblivious positions.

\subsubsection{Exception analysis}
Byeong-Mo et al.~\cite{Byeong-Mo:A_review_on_exception_analysis} gave a comprehensive review on exception analysis.
Magiel Bruntink et al.~\cite{Magiel:Discovering_faults_in_idiom-based_exception_handling} proposed a characterization and evaluation method to statically discover faults in  exception handling.
Fu and Ryder\cite{Fu:Exception-Chain_Analysis} described a static analysis method for exception chains in Java.
Martins et al.~\cite{Alexandre:Testing_Java_Exceptions} presented VerifyEx to test Java exceptions by inserting exceptions at the beginning of try blocks.
Zhang and Elbaum\cite{Zhang:Amplifying_Tests_to_Validate_Exception_Handling_Code} presented an approach that amplifies tests to validate exception handling.
Cornu et al.\cite{Cornu:Exception_Handling_Analysis_and_Transformation_Using_Fault_Injection} proposed a classification of try-catch blocks at testing time.

Those tools rely on test suites to analyze resilience with respect to error-handling. On the contrary, \tripleagent analyzes the system behaviour based on user-level traffic and usages.

\section{Conclusion}\label{sec:conclusion}
In this paper, we have presented \tripleagent, a system which combines automated monitoring, automated perturbation injection and automated resilience improvement. By evaluating \tripleagent on two real-world Java applications, we have shown that it is able to detect weaknesses in exception-handling of Java code and to improve resilience.
In the future, we will further explore the design space of perturbation and failure-obliviousness strategies. For instance, we would like to inject timeout on requests and interactions in asynchronous software.
Our long-term goal is to use \tripleagent in production, and consequently, we will also keep reducing the overhead of \tripleagent at runtime.

\section*{Acknowledgements}
This work was partially supported by the Wallenberg AI, Autonomous Systems and Software Program (WASP) funded by the Knut and Alice Wallenberg Foundation and SSF project TrustFull.

\bibliographystyle{plain}
\bibliography{references}

\begin{thebibliography}{10}

\bibitem{Arcuri:A_practical_guide_for_using_statistical_tests_to_assess_randomized_algorithms_in_software_engineering}
Andrea Arcuri and Lionel~C. Briand.
\newblock A practical guide for using statistical tests to assess randomized
  algorithms in software engineering.
\newblock In {\em Proceedings of the 33rd International Conference on Software
  Engineering, {ICSE} 2011, Waikiki, Honolulu , HI, USA, May 21-28, 2011},
  pages 1--10, 2011.

\bibitem{Basiri:Chaos_Engineering:IEEESoftware}
A.~Basiri, N.~Behnam, R.~de~Rooij, L.~Hochstein, L.~Kosewski, J.~Reynolds, and
  C.~Rosenthal.
\newblock Chaos engineering.
\newblock {\em IEEE Software}, 33(3):35--41, May 2016.

\bibitem{Magiel:Discovering_faults_in_idiom-based_exception_handling}
Magiel Bruntink, Arie van Deursen, and Tom Tourw{\'{e}}.
\newblock Discovering faults in idiom-based exception handling.
\newblock In {\em {ICSE}}, pages 242--251. {ACM}, 2006.

\bibitem{Candea:Crash-Only_Software}
George Candea and Armando Fox.
\newblock Crash-only software.
\newblock In {\em Proceedings of HotOS'03: 9th Workshop on Hot Topics in
  Operating Systems, May 18-21, 2003, Lihue (Kauai), Hawaii, {USA}}, pages
  67--72, 2003.

\bibitem{Byeong-Mo:A_review_on_exception_analysis}
Byeong{-}Mo Chang and Kwanghoon Choi.
\newblock A review on exception analysis.
\newblock {\em Information {\&} Software Technology}, 77:1--16, 2016.

\bibitem{Chang:Chaos_Monkey_Increasing_SDN_Reliability_Through_Systematic_Network_Destruction}
Michael~Alan Chang, Bredan Tschaen, Theophilus Benson, and Laurent Vanbever.
\newblock Chaos monkey: Increasing sdn reliability through systematic network
  destruction.
\newblock {\em SIGCOMM Comput. Commun. Rev.}, 45(4):371--372, August 2015.

\bibitem{Cornu:Exception_Handling_Analysis_and_Transformation_Using_Fault_Injection}
Benoit Cornu, Lionel Seinturier, and Martin Monperrus.
\newblock {Exception Handling Analysis and Transformation Using Fault
  Injection: Study of Resilience Against Unanticipated Exceptions}.
\newblock {\em {Information and Software Technology}}, 57:66--76, January 2015.

\bibitem{Danglot:correctness_attraction}
Benjamin Danglot, Philippe Preux, Benoit Baudry, and Martin Monperrus.
\newblock Correctness attraction: a study of stability of software behavior
  under runtime perturbation.
\newblock {\em Empirical Software Engineering}, 23(4):2086--2119, 2018.

\bibitem{Durieux:Exhaustive_Exploration_of_the_Failure-Oblivious_Computing_Search_Space}
Thomas Durieux, Youssef Hamadi, Zhongxing Yu, Benoit Baudry, and Martin
  Monperrus.
\newblock Exhaustive exploration of the failure-oblivious computing search
  space.
\newblock In {\em 11th {IEEE} International Conference on Software Testing,
  Verification and Validation}, pages 139--149, 2018.

\bibitem{Felipe:An_exploratory_study_on_exception_handling_bugs_in_Java_programs}
Felipe Ebert, Fernando Castor, and Alexander Serebrenik.
\newblock An exploratory study on exception handling bugs in java programs.
\newblock {\em Journal of Systems and Software}, 106:82--101, 2015.

\bibitem{flyvbjerg2006five}
Bent Flyvbjerg.
\newblock Five misunderstandings about case-study research.
\newblock {\em Qualitative inquiry}, 12(2):219--245, 2006.

\bibitem{Fu:Compiler-directed_program-fault_coverage_for_highly_available_Java_internet_services}
C.~Fu, R.~P. Martin, K.~Nagaraja, D.~Wonnacott, T.~D. Nguyen, and B.~G. Ryder.
\newblock Compiler-directed program-fault coverage for highly available java
  internet services.
\newblock In {\em 2003 International Conference on Dependable Systems and
  Networks, 2003. Proceedings.}, pages 595--604, June 2003.

\bibitem{Fu:Exception-Chain_Analysis}
Chen Fu and Barbara~G. Ryder.
\newblock Exception-chain analysis: Revealing exception handling architecture
  in java server applications.
\newblock In {\em Proceedings of the 29th International Conference on Software
  Engineering}, ICSE '07, pages 230--239, Washington, DC, USA, 2007. IEEE
  Computer Society.

\bibitem{Ghosh:Bytecode_Fault_Injection_for_Java_Software}
Sudipto Ghosh and John~L. Kelly.
\newblock Bytecode fault injection for java software.
\newblock {\em J. Syst. Softw.}, 81(11):2034--2043, November 2008.

\bibitem{Gosling:The_Java_Language_Specification_Java_SE_8_Edition}
James Gosling, Bill Joy, Guy~L. Steele, Gilad Bracha, and Alex Buckley.
\newblock {\em The Java Language Specification, Java SE 8 Edition}.
\newblock Addison-Wesley Professional, 1st edition, 2014.

\bibitem{Seungjae:DOCTOR}
Seungjae Han, K.~G. Shin, and H.~A. Rosenberg.
\newblock Doctor: an integrated software fault injection environment for
  distributed real-time systems.
\newblock In {\em Proceedings of 1995 IEEE International Computer Performance
  and Dependability Symposium}, pages 204--213, April 1995.

\bibitem{Huang:Software_Rejuvenation_Analysis_Module_and_Applications}
Yennun Huang, Chandra M.~R. Kintala, Nick Kolettis, and N.~Dudley Fulton.
\newblock Software rejuvenation: Analysis, module and applications.
\newblock In {\em Digest of Papers: FTCS-25, The Twenty-Fifth International
  Symposium on Fault-Tolerant Computing, Pasadena, California, USA, June 27-30,
  1995}, pages 381--390, 1995.

\bibitem{Ghani:FERRARI}
Ghani~A. Kanawati, Nasser~A. Kanawati, and Jacob~A. Abraham.
\newblock {FERRARI:} {A} tool for the validation of system dependability
  properties.
\newblock In {\em {FTCS}}, pages 336--344. {IEEE} Computer Society, 1992.

\bibitem{Kao:A_fault_injection_and_monitoring_environment_for_tracing_the_UNIX_system_behavior_under_faults}
W.~I. Kao, R.~K. Iyer, and D.~Tang.
\newblock Fine: A fault injection and monitoring environment for tracing the
  unix system behavior under faults.
\newblock {\em IEEE Transactions on Software Engineering}, 19(11):1105--1118,
  Nov 1993.

\bibitem{Hyosoon:SFIDA_a_software_implemented_fault_injection_tool_for_distributed_dependable_applications}
Hyosoon Lee, Youngshik Song, and Heonshik Shin.
\newblock Sfida: a software implemented fault injection tool for distributed
  dependable applications.
\newblock In {\em Proceedings Fourth International Conference/Exhibition on
  High Performance Computing in the Asia-Pacific Region}, volume~1, pages
  410--415 vol.1, May 2000.

\bibitem{Madeira:RIFLE}
Henrique Madeira, M{\'{a}}rio~Zenha Rela, Francisco Moreira, and
  Jo{\~{a}}o~Gabriel Silva.
\newblock {RIFLE:} {A} general purpose pin-level fault injector.
\newblock In {\em Dependable Computing - EDCC-1, First European Dependable
  Computing Conference, Berlin, Germany, October 4-6, 1994, Proceedings}, pages
  199--216, 1994.

\bibitem{Alexandre:Testing_Java_Exceptions}
Alexandre~Locci Martins, Simone Hanazumi, and Ana Cristina~Vieira de~Melo.
\newblock Testing java exceptions: An instrumentation technique.
\newblock In {\em {COMPSAC} Workshops}, pages 626--631. {IEEE} Computer
  Society, 2014.

\bibitem{Monperrus:Automatic_Software_Repair_A_Bibliography}
Martin Monperrus.
\newblock Automatic software repair: {A} bibliography.
\newblock {\em {ACM} Comput. Surv.}, 51(1):17:1--17:24, 2018.

\bibitem{Natella:Assessing_Dependability_with_Software_Fault_Injection_A_Survey}
Roberto Natella, Domenico Cotroneo, and Henrique~S. Madeira.
\newblock Assessing dependability with software fault injection: A survey.
\newblock {\em ACM Comput. Surv.}, 48(3):44:1--44:55, February 2016.

\bibitem{Parasyris:GemFI_A_Fault_Injection_Tool_for_Studying_the_Behavior_of_Applications_on_Unreliable_Substrates:DSN}
Konstantinos Parasyris, Georgios Tziantzoulis, Christos~D. Antonopoulos, and
  Nikolaos Bellas.
\newblock Gemfi: {A} fault injection tool for studying the behavior of
  applications on unreliable substrates.
\newblock In {\em 44th Annual {IEEE/IFIP} International Conference on
  Dependable Systems and Networks, {DSN} 2014, Atlanta, GA, USA, June 23-26,
  2014}, pages 622--629, 2014.

\bibitem{Perkins:Automatically_patching_errors_in_deployed_software}
Jeff~H. Perkins, Sunghun Kim, Samuel Larsen, Saman~P. Amarasinghe, Jonathan
  Bachrach, Michael Carbin, Carlos Pacheco, Frank Sherwood, Stelios Sidiroglou,
  Greg Sullivan, Weng{-}Fai Wong, Yoav Zibin, Michael~D. Ernst, and Martin~C.
  Rinard.
\newblock Automatically patching errors in deployed software.
\newblock In {\em Proceedings of the 22nd {ACM} Symposium on Operating Systems
  Principles 2009, {SOSP} 2009, Big Sky, Montana, USA, October 11-14, 2009},
  pages 87--102, 2009.

\bibitem{Qin:Rx_Treating_bugs_as_allergies_a_safe_method_to_survive_software_failures}
Feng Qin, Joseph Tucek, Yuanyuan Zhou, and Jagadeesan Sundaresan.
\newblock Rx: Treating bugs as allergies - a safe method to survive software
  failures.
\newblock {\em {ACM} Trans. Comput. Syst.}, 25(3):7, 2007.

\bibitem{Rigger:Introspection_for_C_and_its_Applications_to_Library_Robustness}
Manuel Rigger, Rene Mayrhofer, Roland Schatz, Matthias Grimmer, and Hanspeter
  M{\"{o}}ssenb{\"{o}}ck.
\newblock Introspection for {C} and its applications to library robustness.
\newblock {\em Programming Journal}, 2(2):4, 2018.

\bibitem{Rigger:Lenient_Execution_of_C_on_a_Java_Virtual_Machine}
Manuel Rigger, Roland Schatz, Matthias Grimmer, and Hanspeter
  M{\"{o}}ssenb{\"{o}}ck.
\newblock Lenient execution of {C} on a java virtual machine: or: How {I}
  learned to stop worrying and run the code.
\newblock In {\em Proceedings of the 14th International Conference on Managed
  Languages and Runtimes, ManLang 2017, Prague, Czech Republic, September 27 -
  29, 2017}, pages 35--47, 2017.

\bibitem{rinard2004enhancing}
M.~Rinard, C.~Cadar, D.~Dumitran, D.M. Roy, T.~Leu, and W.S. Beebee~Jr.
\newblock Enhancing server availability and security through failure-oblivious
  computing.
\newblock In {\em Proceedings of the 6th conference on Symposium on Operating
  Systems Design \& Implementation}, pages 21--21. USENIX Association, 2004.

\bibitem{Sidiroglou:ASSURE}
Stelios Sidiroglou, Oren Laadan, Carlos Perez, Nicolas Viennot, Jason Nieh, and
  Angelos~D. Keromytis.
\newblock {ASSURE:} automatic software self-healing using rescue points.
\newblock In {\em Proceedings of the 14th International Conference on
  Architectural Support for Programming Languages and Operating Systems,
  {ASPLOS} 2009, Washington, DC, USA, March 7-11, 2009}, pages 37--48, 2009.

\bibitem{Silva:A_view_on_the_past_and_future_of_fault_injection_2013}
Nuno Silva, Ricardo Barbosa, Jo{\~{a}}o~Carlos Cunha, and Marco Vieira.
\newblock A view on the past and future of fault injection.
\newblock In {\em 2013 43rd Annual {IEEE/IFIP} International Conference on
  Dependable Systems and Networks (DSN), Budapest, Hungary, June 24-27, 2013},
  pages 1--2, 2013.

\bibitem{Sullivan:Software_Defects_and_their_Impact_on_System_Availability}
Mark Sullivan and Ram Chillarege.
\newblock Software defects and their impact on system availability: {A} study
  of field failures in operating systems.
\newblock In {\em Proceedings of the 1991 International Symposium on
  Fault-Tolerant Computing, Montreal, Canada}, pages 2--9, 1991.

\bibitem{Kouwe:HSFI_Accurate_Fault_Injection_Scalable_to_Large_Code_Bases:DSN}
Erik van~der Kouwe and Andrew~S. Tanenbaum.
\newblock {HSFI:} accurate fault injection scalable to large code bases.
\newblock In {\em 46th Annual {IEEE/IFIP} International Conference on
  Dependable Systems and Networks, {DSN} 2016, Toulouse, France, June 28 - July
  1, 2016}, pages 144--155, 2016.

\bibitem{Wei:Quantifying_the_Accuracy_of_High-Level_Fault_Injection_Techniques_for_Hardware_Faults:DSN}
Jiesheng Wei, Anna Thomas, Guanpeng Li, and Karthik Pattabiraman.
\newblock Quantifying the accuracy of high-level fault injection techniques for
  hardware faults.
\newblock In {\em 44th Annual {IEEE/IFIP} International Conference on
  Dependable Systems and Networks, {DSN} 2014, Atlanta, GA, USA, June 23-26,
  2014}, pages 375--382, 2014.

\bibitem{DingYuan:Simple_Testing_Can_Prevent_Most_Critical_Failures}
Ding Yuan, Yu~Luo, Xin Zhuang, Guilherme~Renna Rodrigues, Xu~Zhao, Yongle
  Zhang, Pranay Jain, and Michael Stumm.
\newblock Simple testing can prevent most critical failures: An analysis of
  production failures in distributed data-intensive systems.
\newblock In {\em 11th {USENIX} Symposium on Operating Systems Design and
  Implementation}, pages 249--265, 2014.

\bibitem{Zhang:Amplifying_Tests_to_Validate_Exception_Handling_Code}
Pingyu Zhang and Sebastian Elbaum.
\newblock Amplifying tests to validate exception handling code: An extended
  study in the mobile application domain.
\newblock {\em ACM Trans. Softw. Eng. Methodol.}, 23(4):32:1--32:28, September
  2014.

\bibitem{Ziade:A_Survey_on_Fault_Injection_Techniques}
Haissam Ziade, Rafic~A. Ayoubi, and Raoul Velazco.
\newblock A survey on fault injection techniques.
\newblock {\em Int. Arab J. Inf. Technol.}, 1(2):171--186, 2004.

\end{thebibliography}

\end{document}